\documentclass[]{interact}
\usepackage{amsmath,amssymb,bm}
\newcommand{\braket}[1]{\langle {#1} \rangle }
\newcommand{\ket}[1]{|{#1} \rangle }

\usepackage{amsmath}	
\usepackage{color}
\usepackage{txfonts}
\usepackage{hyperref}
\begin{document}
\title{Probable observation of the  Cooper pair correlation length in superfluid nuclei}
\author{\name{R. A. Broglia\textsuperscript{a}\textsuperscript{b}, F. Barranco\textsuperscript{c}, G. Potel\textsuperscript{d} and E. Vigezzi\textsuperscript{e}} \affil{\textsuperscript{a}The Niels Bohr Institute, University of Copenhagen, DK-2100 Copenhagen, Blegdamsvej 17, Denmark;  \textsuperscript{b}Dipartimento di Fisica, Universit\`a degli Studi di Milano, Via Celoria 16, I-20133 Milano, Italy; \textsuperscript{c}Departamento de F\'isica Aplicada III, Escuela Superior de Ingenieros, Universidad de Sevilla, Camino de los Descubrimientos, Sevilla, Spain; \textsuperscript{d}Lawrence Livermore National Laboratory, Livermore, California 94550, USA; \textsuperscript{e}INFN Sezione di Milano, Via Celoria 16, I-20133 Milano, Italy.}
}

\maketitle
{\abstract{The phenomenon of low-temperature superconductivity observed in many metals, that is, the conduction of current without resistance, is intimately associated with the condensation in a single, quantum coherent state, known as the Bardeen-Cooper-Schrieffer (BCS) state, of weakly bound (meV), very extended, strongly overlapping pairs (quasi-bosons) of electrons (fermions) known as  Cooper pairs (the center of mass of $10^6$ pairs falls within the volume of a single pair) all carrying the same phase.  Systematic experimental studies of the associated  coherence length $\xi\approx10^4$\AA $\; (\gg r_s\approx 1-3$\AA, Wigner-Seitz radius) have been made. Atomic nuclei, made out of charged protons and uncharged neutrons, namely fermions, are at absolute zero temperature in the ground state, which can in many cases (open shell nuclei), be described as a BCS state, a condensate of, e.g., neutron Cooper pairs. From here the connotation of superfluid nuclei. While the extension of the  BCS theory of superconductivity to the atomic nucleus has been successful beyond expectation {\bf--}explaining from the value of the moment of inertia of deformed nuclei to the life time of fission and of exotic decay processes{\bf--}  to our knowledge, no measurement of the nuclear coherence length \mbox{{\bf--}expected to be much larger than nuclear dimensions{\bf--}} has been reported in the literature. Recent studies of effective charged Cooper pair transfer  across a nuclear Josephson-like junction, transiently established in  heavy ion collisions between superfluid nuclei, have likely changed the situation. They have provided the experimental input for a quantitative estimate of the nuclear coherence length. Also the basis for a nuclear analogue of the alternating current (ac) Josephson effect. Namely the fact that a bias $V$ applied to the junction does not determine the intensity of a supercurrent (Ohm's law) circulating through it, but the frequency of Cooper pairs alternating current, of frequency $\nu=2e\times V/h$.}}

\section{Introduction}\label{S1}

In the paper which marks the start of  quantum mechanics as we know it today, Heisenberg writes \cite{Heisenberg:25bis} ``It is well known that the formal rules which are used in quantum theory for calculating observable quantities such as the energy of the hydrogen atom may be seriously criticized on the ground that they contain, as basic element, relationships between quantities that are apparently unobservable in principle, e.g., position and period of revolution of the electron.'' Considerations made after having expressed the aim ``\dots to establish a basis for theoretical quantum mechanics founded exclusively upon relationships between quantities which in principle are observable''.

Within this context, and with over five decades of experience concerning the phenomenon of nuclear BCS  (\cite{Bohr:58}; see also \cite{Broglia:13} and references therein),  one could posit that the   nuclear correlation length $\xi$ (Eq. (\ref{eq:2})), expected to be much larger than nuclear dimensions,  is  not observable.
To assess the validity of the above statement, it seems important to remind Bohr's advocacy to limit ``the use of the word phenomenon to refer exclusively to observations obtained under specific circumstances, including account of the whole experiment''\cite{Schilpp:51}. Namely, in the present case, two-nucleon transfer {\bf --}or better tunneling {\bf --} processes below the Coulomb barrier between superfluid nuclei.
 \section{Cooper pairs}\label{S3}
 As Cooper  showed \cite{Cooper:92}, a pair of electrons moving in time-reversal states $(\nu,\tilde\nu)$ \mbox{$(\equiv(\mathbf k\uparrow,-\mathbf k\downarrow))$} above a non-interacting Fermi sea displays a bound state {\bf--} $\ket{\Phi_0}\sim\sum_{\nu>0}c_\nu P^\dagger_\nu \ket{0}$, where $P_\nu^\dagger =a^\dagger_\nu a^\dagger_{\bar \nu}$ creates, acting in the vacuum state $\ket{0}$, a pair of electrons in time-reversal states{\bf --}  provided the interaction is attractive, no matter how weak it is.

 {\it Condensed matter}-- In a metal, the passage of an electron through a region of he lattice implies a slight increase of negative charge. It results in a slight attraction of ions and eventually a small excess of positive charge which a second electron will experience. Because $M_{ion}\gg m_e$, only electrons far apart from each other ($\approx 10^4$\AA), and thus feeling a very weak Coulomb repulsion, will interact through the exchange of lattice phonons  \cite{Bardeen:92}, \cite{Frohlich:52}. In other words, because  retardation effects (phonon mediated pairing) permit an attraction that is non-local in space and time, the electrons can avoid most of the repulsion \cite{Schrieffer:92b}. That is, one is confronted essentially with weak binding by deconfinement (see Sect. \ref{S5.1}).

 The condensation  of these ($L=0, S=0$) weakly bound, very extended, strongly overlapping quasi-bosonic  entities ($(P^\dagger_\nu)^2=0$, $P_\nu^\dagger P_{\nu'}^\dagger\neq 0$), leads to a transition between the normal  ({\bf N}) and the superconducting phases ({\bf S}).  The many pair wavefunction $\ket{BCS}$ describing this phase displays a probability amplitude $V_\nu$ that the pair of fermions state ($\nu,\tilde \nu$) is occupied, and at the same time, a probability amplitude $U_\nu=\left(1-V_\nu^2\right)^{1/2}$ that it is empty, a property which can be considered the basic BCS {\it{ansatz}}. As a consequence Cooper pairs can scatter and lower their energy.

 The $\ket{BCS}$ state, written for the first time by Schrieffer \cite{Schrieffer:92},  reads
 \begin{align}
   \label{eq:14}
  \ket{BCS}=\prod_{\nu>0}\left(U'_\nu+e^{-2i\phi}V'_\nu P^\dagger_\nu\right)\ket{0}, 
 \end{align}
 where  $U_\nu=U'_\nu$ and   $V_\nu=V'_\nu e^{-2i\phi}$ ($U_\nu',V_\nu'$ real; $c_\nu=e^{-i2\phi}c_\nu',\;c_\nu'=V'_\nu/U'_\nu$) are the BCS \cite{Bardeen:57a,Bardeen:57b} occupation probability amplitudes, while  $\phi$ is the gauge angle.  The  $\ket{BCS}$  state displays an energy  gap  $\Delta$  for both single-pair translation and dissociation (breaking of Cooper pairs), phenomenon known as Off-Diagonal-Long-Range-Order (ODLRO) \cite{Penrose:51,Penrose:56,Anderson:96,Yang:62}.  The  binding energy of a Cooper pair is $\approx2\Delta$ for electrons moving close to the Fermi energy. Consequently $2\Delta/\hbar$ can be viewed as the effective angular frequency for the back and forth $(L=0)$ oscillations describing the intrinsic motion of the Cooper pair partners, the quantity $\hbar \text{v}_{F}/2\Delta$ where $\text{v}_F$ is the Fermi  velocity, providing a measure of the dimension of the Cooper pair. Following Pippard we shall use,
  \begin{align}
   \label{eq:2}
   \xi=\frac{\hbar \text{v}_F}{\pi\Delta'},
 \end{align}
 as the Cooper pair  mean square radius (correlation length) \cite{Schrieffer:64}.

 One can write the pairing gap as $\Delta=\Delta'e^{-2i\phi}=G\alpha_0$, where
 \begin{align}
   \label{eq:7}
   \alpha_0=\braket{BCS|\sum_{\nu>0}P^\dagger_\nu |BCS}=e^{-2i\phi}\alpha_0',
 \end{align}
 $G$ being the pairing coupling constant, 
 \begin{align}
   \label{eq:8}
   \alpha'_0=\sum_{\nu}U'_\nu V'_\nu
 \end{align}
 the number of Cooper pairs, while
 \begin{align}
   \label{eq:9}
   n'_s=\frac{\alpha_0'}{\mathcal V}
 \end{align}
 is the so called abnormal density, $\mathcal V$ being an appropriate volume element \cite{Ginzburg:04}. The quantity (\ref{eq:7}) is the order parameter of the superconducting state, which one also finds at the basis of the fact that pair tunneling  is the specific probe of BCS  Cooper pair condensation. 

 In keeping with the basic BCS {\it ansatz} ($U_\nu V_\nu\neq0$ in an energy range $\approx2\Delta$ around the Fermi energy) and Eqs. (\ref{eq:7}) and (\ref{eq:8}), the $\ket{BCS}$ state  violates pair of particle number conservation. The quantity $\alpha_0'$ provides a quantitative measure of the associated deformation in the two dimensional gauge space which defines a privileged orientation in this space measured by  $\phi$ (angle between the laboratory system $z$-axis and that of the intrinsic body fixed reference frame). Because all gauge orientations of an isolated superconductor have the same energy, $\ket{BCS}$ is a highly degenerate state, which can be viewed as the intrinsic state of rotation in gauge space of the system as a whole, and thus of the associated pairing rotational bands. These elementary modes of excitation have been systematically studied in nuclei in terms of individual quantum states, making use of two-nucleon   transfer tunneling (see e.g. \cite{Bohr:76,Bes:66,Brink:05,Broglia:00,Hinohara:16}; see also \cite{Anderson:66}). In what follows we shall use, for simplicity, the symbol $\Delta$ for both $\Delta$ and $\Delta'$ unless explicitly stated.  
 
{\it Atomic nuclei}-- The nuclear structure exhibits a number of similarities with the electron structure of metals. In both cases, one is dealing with a system of fermions which can be described in first approximation in terms of independent particle motion. Nonetheless, in both systems, important correlations in the single particle motion arise from the action of the forces between the particles, in particular among those moving in time reversal states close to the Fermi energy. Repulsive in the electronic case (Coulomb interaction) marginally overwhelmed by the exchange of phonons, attractive in the nuclear one (strong force; of note however the Coulomb force acting between protons). Furthermore, in the nuclear case the exchange of collective, mainly surface vibrations, contribute about half of the total attractive strength \cite{Barranco:99,Saperstein:12,Avdenkov:12,Lombardo:12,Idini:15,Barranco:05}. A further point of contact with metallic systems, where the lattice phonons provide the designed glue to bind pairs of ($\nu,\tilde\nu$) electrons into Cooper pairs.

Pairing in nuclei was introduced by Bohr, Mottelson and Pines (\cite{Bohr:58}; see also \cite{Bohr:69,Bohr:75,Bohr:76,Mottelson:76}), few months after the publication of the BCS papers \cite{Bardeen:57a,Bardeen:57b}, in terms of the pairing gap in the intrinsic spectrum of quadrupole deformed nuclei and of the associated moments of inertia of rotations of the system as a whole, as well as making reference to the odd-even mass difference \cite{Mayer:48,Mayer:49,Mayer:55}. If the intrinsic structure of these finite, many body systems could be adequately described in terms of single-particle motion, the first quantity should be of the order of 0.1-0.2 MeV while the observed value is 1.0-1.5 MeV. Concerning the moment of inertia $\mathcal I$, one expects a value equal to the rigid moment of inertia $\mathcal I_{rig}$, in keeping with the fact that the single-particle orbitals are solidly anchored to the mean field. Experimentally, $\mathcal I\approx \tfrac{1}{2}\mathcal I_{rig}$ {\bf--} consistent with a spheroid filled with a low viscosity fluid {\bf--} a result which was explained in terms of nuclear BCS \cite{Belyaev:59,Belyaev:13}.
In other words, in nuclei one finds a pairing gap for both the motion of one pair independent of the others (low level density in even-even systems) as well as in the breaking of a pair (odd-even mass difference), i.e. ODLRO.

The odd-even mass difference, namely the larger stability of nuclei with an even number of like nucleons as compared to that of the neighboring odd numbered nuclei, has played an important role in the understanding of pairing in nuclei, and was already recognized in connection with the introduction of the shell model \cite{Mayer:55}.  
A main difference between metals and atomic nuclei concerning BCS Cooper pair condensation, is associated with the role pairing fluctuations, play in these systems \cite{Bohr:64,Hogassen:61,Bjerregaard:66b,Bes:66,Broglia:67}.

Within this context (shell model and pairing vibrations), the texture of superfluidty in nuclei can be studied one step at a time, in terms of individual quantum states and of single pair transfer processes. Because of the large energy gap found in closed shell nuclei\footnote{That is, $^{A}_Z$X$_{N}$ systems of mass number $A=N+Z$ and magic number (2,8,20,28,50,82,126\dots) of neutrons ($N$) and/or protons ($Z$).} between occupied and empty single-particle states, these systems (e.g. $^{208}_{82}$Pb$_{126}$), are normal (non superfluid) nuclei while the ground states of
$^{210}$Pb ($|^{210}_{82}$Pb$_{128}(gs)\rangle$) can be viewed as a (single) Cooper pair or, as the lowest pair addition mode of $^{208}$Pb populated in a two-nucleon transfer process, while the ground state of $^{62}_{28}$Ni$_{34}$ and  of  $^{114}_{50}$Sn$_{64}$ populated in the reaction $^{116}_{50}$Sn$_{66}$+$^{60}_{28}$Ni$_{32}\to$ $^{114}_{50}$Sn$_{64}(gs)$+$^{62}_{28}$Ni$_{34}(gs)$ (Sect. \ref{S6} below), can be described at profit as BCS condensates of 7 (=(64-50)/2) and 3 (=(34-28)/2) neutron Cooper pairs respectively\footnote{7 and 3 are the number of pairs of neutrons outside the neutron closed shells $N=50$ and $N=28$ of $^{114}$Sn and $^{62}$Ni respectively.}, members of pairing rotational bands where the number of particles play the role of angular momentum (see e.g. \cite{Potel:13b} and refs therein). See  also \cite{Bracco:96}.

One can posit that Cooper pairs are at the basis of a variety of elementary modes of excitation. Pairing vibrations for single Cooper pairs, pairing rotations for few ones and supercurrents, for macroscopic amounts of them. Common to all of these  modes when probed with Cooper pair tunneling one finds the dominance of successive transfer (tunneling) of entangled fermion partners.

 \section{Correlation length in superconductors}
 Let us assume microwave  electromagnetic radiation, that is, radiation of  frequency  300 GHz--300 MHz, falls on the surface of a normal metal for which the electronic mean free path $l$ is short ($l<\delta$, see below), so that the dependence of the current density $J$ of carriers of charge $q=e$ (single electrons) on the electric field $\mathbf E$ can be described by the local relation $\mathbf J(\mathbf r)=\sigma  \mathbf E (\mathbf r)$, where $\sigma$ is the conductivity. It is found that propagation inside the metal is attenuated according to $E=E_0\exp[i(kz-\omega t)-z/\delta]$ where $\delta=(2/\sigma\mu_0\omega)^{1/2}$ is the skin penetration depth (of the order of 0.7$\times10^4$\AA$\,$), due to the appearance  of eddy currents produced by the time dependence of the associated magnetic field. This is known as the normal skin effect. The quantity that is usually measured is the surface impedance  $Z_s$ (\cite{Dheer:61}; see also \cite{Tilley:90}), ratio of the electric field at the surface to the total current flowing across a unit line at the surface. 
 In the case in which $\delta<l$ (anomalous skin effect), one has to generalize the local Ohm's law into a non-local equation, allowing for the fact that electrons accelerated by the electric field may travel some distance before they are scattered, and that the current density is not related to the local field, but to an appropriate average over a domain defined by the decay factor $\exp(-\rho/l)$ ($\rho=|\mathbf r-\mathbf r'|$) \cite{Chambers:52}.

 A similar suggestion was made by Pippard \cite{Pippard:53} in generalizing the local London \cite{London:35,London:54} equation for the supercurrent density $\mathbf J_s=-\frac{n_s' e^2}{m_e}\mathbf A$ of carriers of charge $q=2e$ (Cooper pairs), to a non-local one which is able also to describe pure (free from impurities) superconductors, in which case the averaging domain is defined by the decay factor $\exp(-\rho/\xi^{\,'})$ $(1/\xi^{\,'}=1/\xi+1/(\alpha l))$. In the case of tin, the numerical values of $\alpha\approx 0.8$ for the proportionality factor of $l$, and  $\xi\approx 1.2\times 10^4$ \AA$\,$ for the coherence (correlation) length, were estimated from penetration depth measurements, carried out in wires $\approx1.4$ cm (resonance frequency $\approx9400$ MHz) and 0.5 mm of diameter \cite{Pippard:53}. Following Pippard's breakthrough, a number of techniques were developed to measure the penetration length (see e.g. \cite{Mersevey:69} and refs. therein).  This value of $\xi$ was microscopically validated by BCS theory (Eq. 5.50 \cite{Bardeen:57b}; see also Eq. (\ref{eq:2})), and implies that no important variations of the electromagnetic field and thus neither of $n_s'$ takes place within a  range $\xi$.

 Summing up, London's equation  leads to $H=H_0\exp(-z/\lambda_L$), where $\lambda_L=(m_ec^2/(4\pi n_s'e^2))^{1/2}$ is known as London penetration depth. A quantity of the order of few hundreds of \AA. The values of $\lambda$ emerging from Pippard's equation are considerably larger than $\lambda_L$ in overall agreement with the data, aside from displaying the right dependence on $l$, quantity which changes from the value of $10^6$\AA$\,$ in the purest available tin \cite{Pippard:53}, to less than $10^2$\AA$\,$ when doped with 3\% indium.
  \section{The Josephson effect}
 In the case of two superconductors (S) separated by an insulating barrier of thickness $d$ of few nanometers, namely much smaller than $\xi$,  single electrons can traverse the barrier, although with rather small probability ($P_1\approx10^{-10}$) \cite{Pippard:12}. One thus talks about  a weak link. They were employed by Giaever to carry out tunneling experiments of single-electron carriers (in both N-S and S-S links), which were instrumental in the determination of the pairing gap \cite{Giaver:73}.

 If a single electron has such a small probability to get through the barrier, the simultaneous tunneling of two electrons (probability $P_1^2$) will not be observed. To this Josephson argued that the wavefunctions of the electrons in the pair are (gauge) phase coherent. One has to add the amplitudes before taking modulus squared (see also \cite{Anderson:64b}). It is like interference in optics with phase coherent wave mixing, and the probability of  pair tunneling is comparable to the probability for a single electron. After the Cooper pair tunneling effects predicted by Josephson \cite{Josephson:62,Josephson:73} were confirmed \cite{Anderson:63,Shapiro:63}, the weak link set up became known as a Josephson junction.

 The  main effects predicted by Josephson were: {(\bf{a}) unbiased junction} (direct current (dc) Josephson effect); the small but finite overlap of the BCS condensates described by $\ket{BCS (\ell)}$ and $\ket{BCS (r)}$ is sufficient to lock the associated gauge phase difference ($\delta\phi(\mathbf R)=\phi_{\ell}-\phi_{r}$), a function which plays the role of the velocity potential leading to a momentum which acts on the center of mass ($\mathbf R$) of  Cooper pairs. The associated zero-voltage direct current\footnote{It is of notice that one should use a gauge invariant phase, see for example \cite{Anderson:64b}.} $(I=I_c\sin(\delta\phi))$ of carriers of charge $q=2e$ (Cooper pairs) and maximum (critical) value $I_c=\frac{\pi}{e}\frac{\Delta_\ell\Delta_r}{\Delta_\ell+\Delta_r}\frac{1}{R_b}$ is undamped, because the internal degrees of freedom of the Cooper pairs are frozen by the pairing gap, $R_b$ being the junction resistance to N-currents ($q=e$).   In the above relation $\Delta_\ell$ and $\Delta_r$ are the pairing gaps of the left ($\ell$) and right ($r$) superconductors with respect to the junction.  Similarly, concerning the gauge angles $\phi_\ell$ and $\phi_r$; {\bf (b) biased junction} (alternating current (ac));  when there is a  constant voltage $V$ applied across the junction, circulation of an alternating supercurrent $I=I_c\sin(\nu_J t)$ of Cooper pairs with critical value $I_c$ and Josephson frequency $\nu_J=V(2e)/h$ was predicted. Because the frequency is so high (THz), the $\sin(\nu_J t)$ function averages out to essentially zero. In keeping with the fact that the  junction is biased, each time a Cooper pair tunnels from one side of the barrier to the other, there is an energy difference $\Delta E=V\times 2e$. Being the process free of dissipation (alternating supercurrent), to leave the quasiparticle distribution unchanged, Cooper pairs can tunnel with emission of photons of frequency $\nu_J$ as has been experimentally observed both in absorption \cite{Shapiro:63} and in emission  \cite{Lindelof:81} (see also \cite{Rogalla:12}).
 \subsection{Critical current}
 In a supercurrent, the momentum $\mathbf q$ acting on the center of mass of the Cooper pair gives rise to a violation of time reversal invariance $((\mathbf k+\mathbf q)\uparrow,(-\mathbf k+\mathbf q)\downarrow )$. Violation which, for $q_c=1/\xi$,  becomes critical (depairing), leading to breaking of  Cooper pairs and thus,  to a phase transition,  from the superconducting  to the normal  phases. This is in keeping with the fact that the momentum $\mathbf q$ shifts the energy of the quasiparticle of momentum $\mathbf k (E_{\mathbf k})$ by $\hbar \text{v}_F q$. Consequently, for $q=q_c$, $(\hbar \text{v}_F)\times (1/\xi)=\pi\Delta$. That is, a value corresponding to approximately the depairing energy, and another possible way to read the value of $\xi$ from  condensed matter experiments (see also below, critical current).  
 A similar result is obtained by considering Josephson's critical current, which we write for simplicity for two equal superconductors (i.e. $\Delta_\ell=\Delta_r=\Delta$), as
 \begin{align}
   \label{eq:3}
   I_c=\frac{\pi}{4}\frac{V_{eq}}{R_b},
 \end{align}
 where
 \begin{align}
   \label{eq:4}
   V_{eq}=\frac{2\Delta}{e},
 \end{align}
 is the equivalent (Giaever's) depairing potential leading to a transition between the S-S to the {\bf S-Q} tunneling regimes,  Q referring  to quasiparticles. In other words, a transition between a ground-ground state supercurrent of Cooper pairs, to a current of single electrons  involving gs-gs and gs-quasiparticle transitions.  Equation (\ref{eq:3}) can then be written as
 \begin{align}
   \label{eq:5}
   I_c=\frac{\pi}{4}I_N,
 \end{align}
 implying that, within a factor $(\pi/4)$
 \begin{align}
   \label{eq:6}
   P_2\approx P_1,
 \end{align}
 where $P_1$ is the tunneling probability of normal (N), single electron carriers current, while $P_2$ is that of a Cooper pair supercurrent. The above results,  which melts Josephson's (with non-negligible contributions from Anderson \cite{Anderson:64b}) and Giaever's \cite{Giaver:73},  theoretical predictions and experimental work respectively, is in keeping with the fact that, as already stated above, in the calculation of $P_2$ one has to add the phased amplitudes of each partner electron, before taking the modulus squared.
 \section{Entanglement}
 Schr\"odinger, one of the originators of quantum mechanics states \cite{Schrodinger:35} that not {\it one} but {\it the} characteristic trait of quantum mechanics is entanglement. That is, the phenomenon resulting from the fact that ``when two systems, of which we know the states by their respective representatives, enter in temporary physical interaction due to known forces between them, and when after a time of mutual influence the systems separate again, then they can no longer be described in the same way as before, viz. by endowing each of them with a representative of its own\dots By the interaction the two representatives (or $\Psi$-function) have become entangled''.

 Below the critical temperature, pairs of electrons moving in time reversal states close to the Fermi surface become, through the exchange of lattice phonons,  entangled over a correlation length of the order of $10^{4}$\AA,         giving rise to Cooper pairs which eventually condense into the coherent $\ket{BCS}$ state. As a consequence, Cooper pair tunneling across a Josephson junction can take place one electron at a time, the partner electrons being similarly pairing entangled when each of them is in a different side of the junction $(\xi\gg d)$, than when both are within the same superconductor. Entanglement,  the characteristic trait of quantum mechanics, is  at the basis of the equality $(\ref{eq:6})$.

\subsection{Generalized quantality parameter}\label{S5.1}
The above statement  can be quantitatively expressed through the generalized quantality parameter (\cite{Mottelson:02,Potel:17}; see also \cite{deBoer:48b,deBoer:48,deBoer:57,Nosanow:76}). Namely, the ratio of the quantum fluctuations associated with the  kinetic energy of confinement $\hbar^2/(2m_e\xi^2)\approx4\times10^{-8}$ meV, and the Cooper pair binding energy $\approx 2\Delta\approx3$ meV (the numerial values of $\xi$ and $\Delta$ corresponding to lead), leading to $q_\xi\approx10^{-8}$. In keeping with the fact that potential energy always prefer spatial particle arrangements (generalized rigidity) and fluctuations favor symmetry, that is delocalization \cite{Anderson:84}, the above value of $q_\xi$ implies the presence of conspicuous entanglement between the fermionic partners of a Cooper pair. More precisely, being the quantal kinetic energy of confinement so small, essentially any attractive interaction no matter how weak it is leads to the tethering of one partner to the other. An example of ``entanglement by delocalization''. This is the physical reason at the basis of the fact that  successive is the dominant mechanism in Cooper pair transfer (tunneling) between two weakly coupled superconductors (superfluids) in both condensed matter and in nuclei.  Also of the fact that this reaction process is the specific probe of Cooper pair coherence length, providing quantitative information on the conjecture that, in nuclei, $\xi>R$.

\section{Nuclear correlation length: the experiment}\label{S6}
In what follows we aim at discussing  the possibility of measuring the nuclear Cooper pair mean square radius (correlation length). A possibility brought into focus by a recent breakthrough in the subject of nuclear superfluidity. It was accomplished through the experimental study and the theoretical analysis of one- and two-neutron transfer reactions between two superconducting nuclei (see next Section),  enabled by the use  of magnetic and $\gamma$-ray spectrometers, \cite{Montanari:14} \cite{Montanari:16}  
 \begin{align}
   \label{eq:1}
 ^{116}\text{Sn}+^{60}\text{Ni}\to\left\{
   \begin{array}{l}
     ^{115}\text{Sn}+^{61}\text{Ni}\quad (Q_{1n}=-1.74\text{ MeV}),\quad \text{(a)}\\
     ^{114}\text{Sn}+^{62}\text{Ni}\quad (Q_{2n}=1.307\text{ MeV}).\quad  \text{(b)}
   \end{array}\right.
 \end{align}
 The reactions were carried out at twelve bombarding energies in the range 140.60 MeV $\leq E_{cm}\leq167.95$ MeV. That is, from energies above the Coulomb barrier ($E_B=157.60$ MeV), to well below it. Absolute differential cross sections were measured at $\theta_{cm}=140^\circ$ and transfer probabilities $P_1$ and $P_2$, for one- and two-particle transfer reactions, extracted. Both transfer channels \ref{eq:1} (a) and \ref{eq:1} (b) are inclusive, in keeping with the fact that the beam energy resolution is 2 MeV. It implies that, within this energy range, individual levels populated in the transfer process cannot, in principle, be distinguished and an incoherent sum of their contribution to the transfer cross section, is to be made.

 Gamma-coincidence experiments\footnote{In this connection a $\gamma$-particle coincidence experiment for the system $^{60}$Ni +$^{116}$Sn was carried out \cite{Montanari:16}, and Doppler-corrected $\gamma$-spectra (discrete lines) for $^{60,61,62}$Ni and $^{116,115,114}$Sn in the energy interval 0--1600 keV  displayed in Fig. 2 of that reference. It is of note that these $\gamma$-ray lines can hardly have  relation to  the $\gamma$-strength function displayed in Fig. \ref{fig:4} below.} \cite{Montanari:16}  testify to the fact  that in the case of \ref{eq:1} (b), less than 24\% of the cross section goes to excited states.  Theoretical studies indicate that while channel \ref{eq:1} (b) is dominated by the ground-ground state transition, channel \ref{eq:1} (a) receives incoherent contributions from a number of excited states (see below). 
 \subsection{Nuclear effective charges}\label{S6.1}
If one shines a beam of $\gamma$-rays on a nucleus, the photons can set the center of mass (CM) of the system into oscillation (Thomson scattering), or be absorbed by the system, in which case the center of mass does not oscillate, as a result of the action of the photons dipole field acting in one direction on the protons, and the reaction of the neutrons, mediated by the  proton-neutron strong force, in the opposite direction. This last is a picture that holds also for nuclear $\gamma$-emission.

Writing the dipole moment of the nucleons referred to the CM (intrinsic) system one obtains $\mathbf d=e\sum_{i=1}^A \mathbf r_i-\mathbf R$,where $e$ is the proton charge, and $\mathbf R=\sum_{i=1}^A \mathbf r_i$ is the CM coordinate. Substituted in $\mathbf d$ leads to $\mathbf d=\sum_{p=1}^Z e^{eff}_p \mathbf r_p+\sum_{n=1}^N e^{eff}_n \mathbf r_n$, where $e^{eff}_p=e(1-Z/A)=e N/A$ can be viewed as the proton effective charge and $e^{eff}_n=e(0-Z/A)=-e Z/A$ as the neutron one. In the process (\ref{eq:1}) (b) the effective charge of each transferred neutron is $-e\times\frac{Z_a+Z_A}{A_a+A_A}=-e(78/176)=-e\times 0.443$. Consequently, from this point of view one can talk about superconducting nuclei, and alternating currents of single Cooper pairs.

Within the context of the reaction (\ref{eq:1}) (b),  and making use of relations (\ref{eq:7})-(\ref{eq:9})), we note that $\alpha'_0$($^{62}$Ni$)\approx3$ and $\alpha'_0$($^{116}$Sn$)\approx8$ while $\mathcal V$($^{62}$Ni)$\approx$ 430 fm$^3$ and $\mathcal V$($^{116}$Sn$)\approx$ 860 fm$^3$. Using average values one obtains $n_s'\approx6/650$ fm$^{-3}\approx10^{-2}$ fm$^{-3}$, the nuclear saturation density being 0.17 fm$^{-3}$. Particle number fluctuation associated with the reacting nuclei is thus equal to $\delta N_{rms}=\sqrt{2\alpha_0'}$, and thus 2.5 and 4.0 respectively, implying  fluctuations of the associated gauge angles.

 \section{Results of the experiment}
       \begin{figure}
	\centerline{\includegraphics*[width=10cm,angle=0]{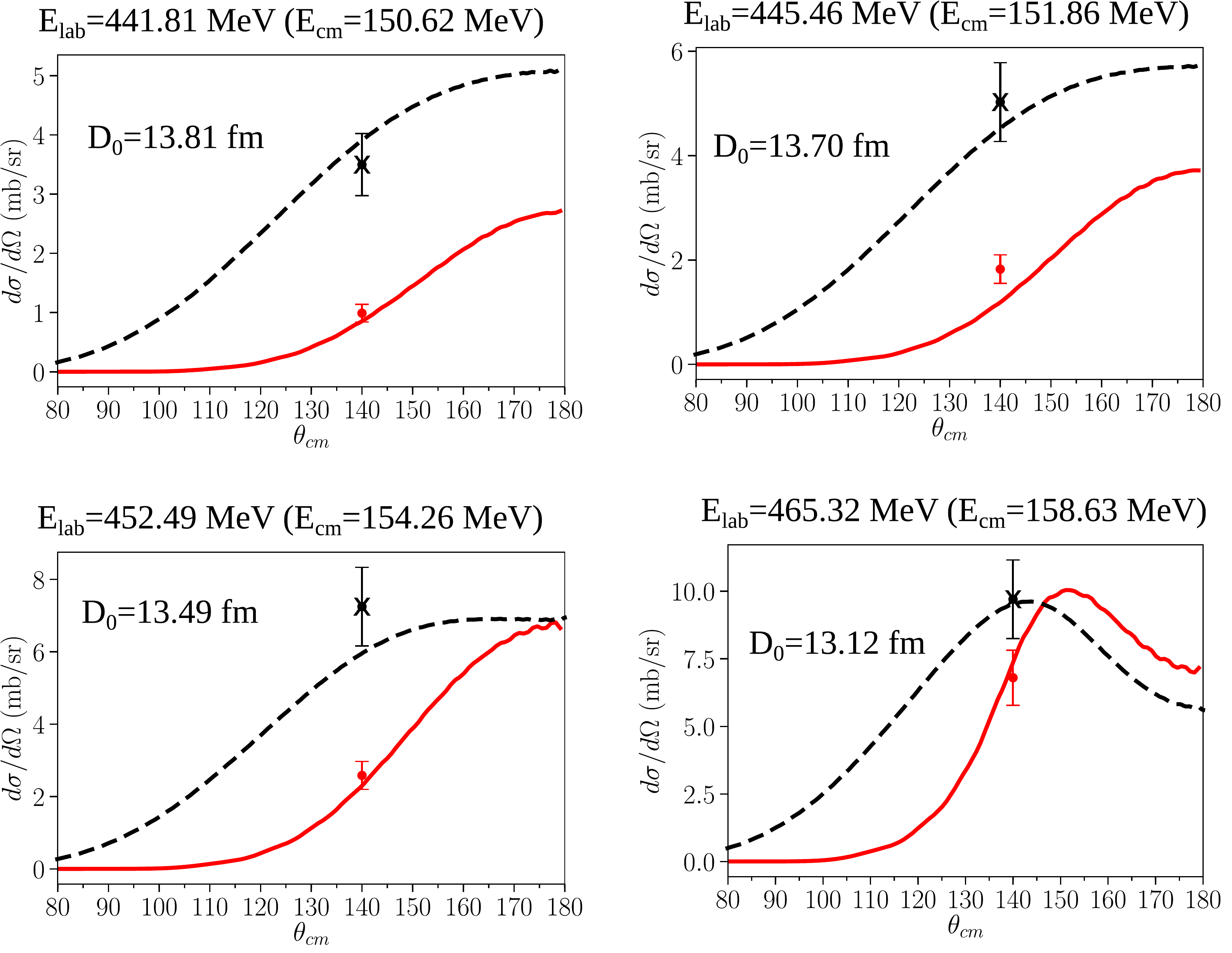}}
	\caption{Absolute differential cross sections associated with the reactions $^{116}$Sn+$^{60}$Ni$\to^{114}$Sn+$^{62}$Ni (continuous line) and $^{116}$Sn+$^{60}$Ni$\to^{115}$Sn+$^{61}$Ni, (dashed line) in comparison with the experimental  data \cite{Montanari:14}: $d\sigma_{2n} /d\Omega|_{\theta_{cm}=140^\circ}$ (solid dots), $d\sigma_{1n} /d\Omega|_{\theta_{cm}=140^\circ}$ (crosses), for the four bombarding energies within the interval 150.62 MeV$\leq E_{cm}\leq 158.63$ MeV, corresponding to the distances of closest approach $D_0$ associated with the first four  points displayed  in Fig. \ref{fig:2}.
	}\label{fig:2x}
  \end{figure}
  \begin{figure}
	\centerline{\includegraphics*[width=9cm,angle=0]{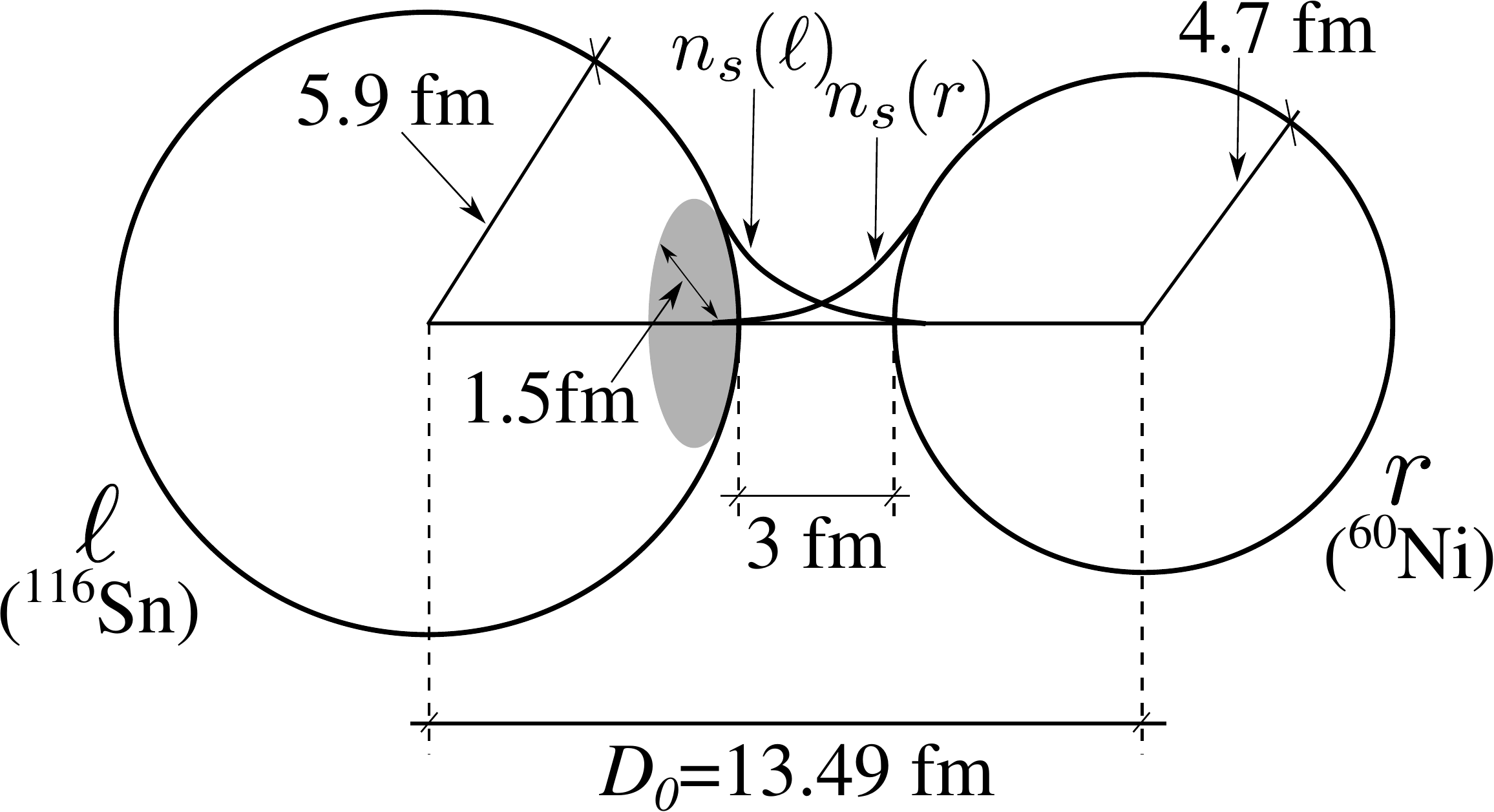}}
	\caption{Schematic representation of the abnormal density overlap of the superconducting nuclei $(\ell)$ $^{116}$Sn and $(r)$ $ ^{60}$Ni at the distance of closest approach $D_0\approx13.5$ fm, which defines the radius of the junction ($\approx1.5$ fm, $S=\pi(1.5\text{ fm})^2\approx 7$ fm$^2$), as well as its length ($L\approx 3$ fm; see \cite{Magierski:17},\cite{Magierski:17b}). A simple estimate of the junction energy is provided by the relation $E_J=(S/L)\times (\hbar^2/2m)n_s'\sin^2(\delta\phi/2)$. Making use of the values of $S$ and $L$ estimated in this figure, and of $\langle e^{i2\delta\phi}\rangle\approx1$ (text), one obtains $E_J\approx0.5$ MeV, see  \cite{Magierski:17b}.}\label{fig:1x}
\end{figure}
One can view the heavy ion reactions (\ref{eq:1}) as a transient nuclear Giaever-like ((a)), and a transient (ac) nuclear Josephson-like ((b)) junctions between two superconducting nuclei, respectively. To calculate the absolute differential cross sections $d\sigma_{1n}/d\Omega|_{\theta_{cm}=140^\circ}$ associated with the one quasiparticle transfer process (\ref{eq:1}) (a), the occupation/emptiness amplitudes ($V_j/U_j$) of the single quasiparticle levels of Sn/Ni are needed.  In the case of Sn they were obtained from  BCS calculations \cite{Montanari:14}, while in the case of Ni from $(d,p)$-data \cite{Lee:09}. Making use of the distorted wave Born approximation (DWBA; see e.g. \cite{Satchler:80} and references therein), and employing the microscopically calculated \cite{Broglia:81b,Pollarolo:83,Broglia:04a}; see also \cite{Sorensen:92b} optical potential (hevy ion reaction dielectric function) reported in  \cite{Montanari:14}, an overall account of the experimental data is obtained \cite{Potel:21} (see Fig. \ref{fig:2x}), by including the incoherent contribution of all (eleven) quasiparticle (Q) states of $^{61}$Ni with energies $\lesssim 2.640$ MeV. A value which is consistent with twice the value of the pairing gap of Ni, and with the S-Q mechanism at the basis of Giaever's single-electron tunneling experiments.

 Concerning the calculation of the absolute differential cross section $d\sigma_{2n}/d\Omega|_{\theta_{cm}=140^\circ}$ associated with (gs-gs) Cooper pair transfer, the corresponding spectroscopic amplitudes $(B_j=((2j+1)/2)^{1/2}U'_j V'_j\times e^{-2i\phi},\; U'_j V'_j$ being BCS coherence factors) were used to calculate, in second order DWBA, the $T$-matrix associated with simultaneous, non-orthogonality and successive transfer. This last one being the overwhelming contribution to the process, in analogy with Josephson's pair tunneling mechanism and Eq. (\ref{eq:6}).

 The role the gauge angle plays in the reaction \ref{eq:1} (b) has been discussed in \cite{Broglia:21} (within this context see also \cite{Magierski:21}). Here we limit ourselves to note that, in keeping with the indeterminacy relations $\delta N\delta\phi\geq1$ one obtains, making use of the estimates made at the end of Sect. \ref{S6.1}, $\delta\phi\approx((1/2.5)^2+(1/4)^2)^{1/2}$ rad$\approx0.5$ rad.  Thus $\langle e^{i2\delta\phi}\rangle\approx\int_0^{0.5}e^{i2\delta\phi} d(\delta\phi)/\int_0^{0.5}d(\delta\phi)\approx\frac{1}{i}(-\frac{1}{2}+i\frac{\sqrt{3}}{2})$, and  $|\langle e^{i2\delta\phi}\rangle|^2\approx1$.

 Within this context, it is also of note  the weak coupling nature of the nuclear link considered (small overlap between abnormal densities and short collision time) for the selected bombarding conditions \cite{Potel:21}  (Fig. \ref{fig:2x}) and thus values of the distance of closest approach $D_0$, as also testified by the value of the junction energy  $E_J\approx0.5$ MeV (see Eq. (1) of \cite{Magierski:17}), obtained making use of the estimates displayed in  Fig.  \ref{fig:1x}. A value rather close to that expected from the Josephson weak link interaction expression \cite{Josephson:62,Anderson:64b} $\Delta_\ell \Delta_r/(\Delta_r+\Delta_\ell)\approx\Delta/2\approx0.7$ MeV, $\Delta\approx1.4$ MeV being a sensible estimate for both Sn and Ni superfluid nuclei.

 Once the consistency between theory and the experimental findings have been assessed with a positive outcome, we try to extract from the data the translation of the relation (\ref{eq:5}), and thus of the critical momentum  $q_c=1/\xi$ (i.e. the correlation length $\xi$), to the nuclear case. To do so, one is first reminded of the fact that in a heavy ion reaction like \ref{eq:1} (b), it is more natural to adjust the critical width of the transient junction (barrier) through which the Cooper pair has to tunnel,that is the critical distance of closest approach $D_0^{(c)}\approx1/q_c\approx\xi$, than the (center of mass) momentum with which it tunnels. The fact that both requirements are, to a large extent physically equivalent, is in keeping with the fact that $k$ and $x$ are conjugated variables.

 Summing up, using  the microscopically calculated optical potential (heavy ion reaction dielectric function), theory  provides  an overall account of the data, in this case for the single Cooper pair transfer process  (Fig. \ref{fig:2x}, see also \cite{Montanari:14} and \cite{Potel:21}).
 In the case of the reaction (Eq. \ref{eq:1} (a)) one is in presence of a S-Q like transfer, while concerning the  (gs-gs) reaction given in Eq. (\ref{eq:1} (b)), one is confronted with a S-S one. Within this context, it is of note that the corresponding values of the real and imaginary components of the microscopic optical potential used in the calculations (see \cite{Montanari:14}, see also p. 111 \cite{Broglia:04a}) are, for the relative distance of closest approach $r=D_0\approx13.5$ fm, -0.9 MeV and 0.6 MeV respectively, namely rather small.

 \subsection{Nuclear equivalent of $I_c$}\label{SA}
    \begin{figure}
	\centerline{\includegraphics*[width=9cm,angle=0]{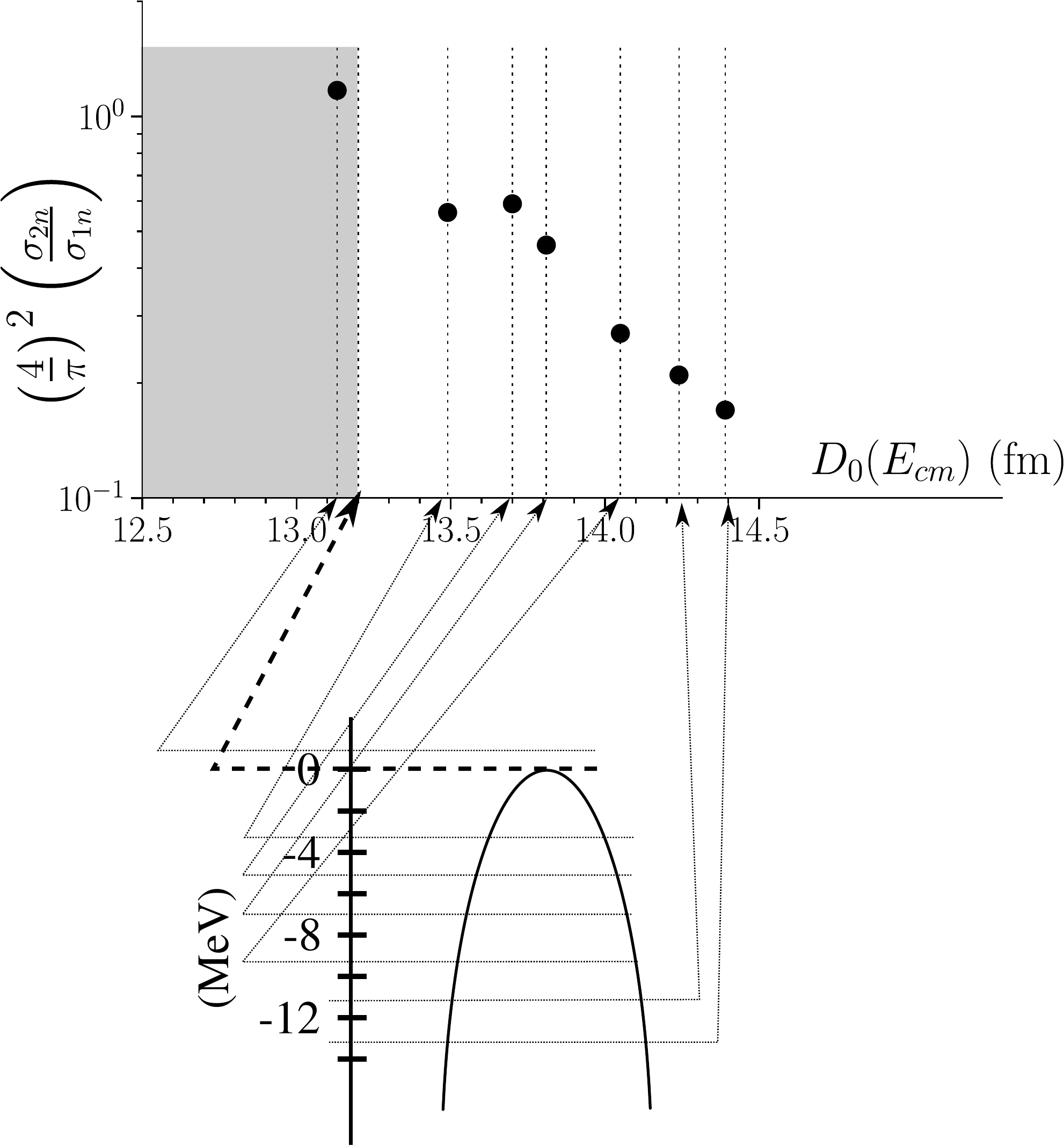}}
	\caption{The ratio $(\sigma_{2n}/\sigma_{1n})$ associated with the reactions $^{116}$Sn$+^{60}$Ni$\to^{114}$Sn$+^{62}$Ni (gs$\to$gs (S-S) process), and $^{116}$Sn$+^{60}$Ni$\to^{115}$Sn$+^{61}$Ni ((S-Q) process) multiplied by the factor $\left(\frac{4}{\pi}\right)^2$, are represented by solid dots  as a function of the distance of closest approach $D_0$ associated with the center of mass bombarding energy. From $E_{cm}=158.63$ MeV ($D_0=13.12$ fm), just 1 MeV above the Coulomb barrier ($E_B=157.60$ MeV, $D_0(E_B)=13.2$ fm, gray zone),  to $E_{cm}=145.02$ MeV ($D_0=14.39$ fm), 12.58 MeV below the Coulomb barrier. In the lower part of the figure, a schematic representation of the Coulomb barrier is given. The difference $E_{cm}-E_B$ (MeV) is plotted along the $y$-axis. Indicated with thin (thick dashed) horizontal lines are the values corresponding to the seven chosen bombarding energies $E_{cm}$ (Coulomb barrier $E_B$). They are prolonged with an arrowed line, to indicate the corresponding $D_0$-values shown in the $x$-axis of the upper plot.}\label{fig:2}
  \end{figure}
 The nuclear equivalent of the relation $I_c/I_N\approx\frac{\pi}{4}$ $(q=1/\xi)$ is expected to be $\sigma_{2n}/\sigma_{1n}\approx\left(\frac{\pi}{4}\right)^2$, ($D_0^{(c)}\approx1/q_c\approx\xi$). 
 As seen from Fig. \ref{fig:2} the quantity $\left(\frac{\pi}{4}\right)^2\left(\frac{\sigma_{2n}}{\sigma_{1n}}\right)$ displays, as a function of the distance of closest approach, a sort of ``plateau'' within the range 13.4 fm$\leq D_0 \leq 13.7$ fm around a value $\approx 0.6$. Although, in hindsight, a better statistics in this region would have been desirable,  we have  chosen the value of 13.5 fm $(D_0=13.49\text{ fm}, E_{cm}=154.26\text{ MeV})$ as representative for that of the correlation length.

  Because the pairing gap associated with the spectroscopic amplitudes $B_j$ is\footnote{Using the $B_j$-spectroscopic amplitudes for Sn, the pairing gap $\Delta=G\sum_j((2j+1)/2)^{1/2}B_j=G\sum_j((2j+1)/2)U_jV_j$ was calculated, leading to $\Delta=1.4$ MeV, and thus to $\xi=13.6$ fm ($\text{v}_F/c\approx0.3$).} $\Delta\approx1.4$ MeV --and thus $\xi\approx13.6$ fm-- one can posit that the data of \cite{Montanari:14} seems to provide, through the specific probe of Cooper pair transfer, an answer to the question, what the nuclear correlation length is.

  \section{The prediction}\label{S7}
 Within this context, and in keeping with the fact that the laboratory bombarding energy associated with $D_0\approx13.5$ fm is $\approx$3.9 MeV/$A$ ($\approx 452.49$ MeV/116), that is an order of magnitude smaller than the Fermi energy, one can expect that there can be time for the nuclear Cooper pair to be transferred back and forth more than once between target and projectile. That is, for more than one cycle of the quasielastic process
 \begin{align}
   \label{eq:12}
   ^{116}\text{Sn}+^{60}\text{Ni}\to\, ^{114}\text{Sn}+^{62}\text{Ni}\to\,^{116}\text{Sn}+^{60}\text{Ni}.
 \end{align}
        \begin{figure}
	\centerline{\includegraphics*[width=10cm,angle=0]{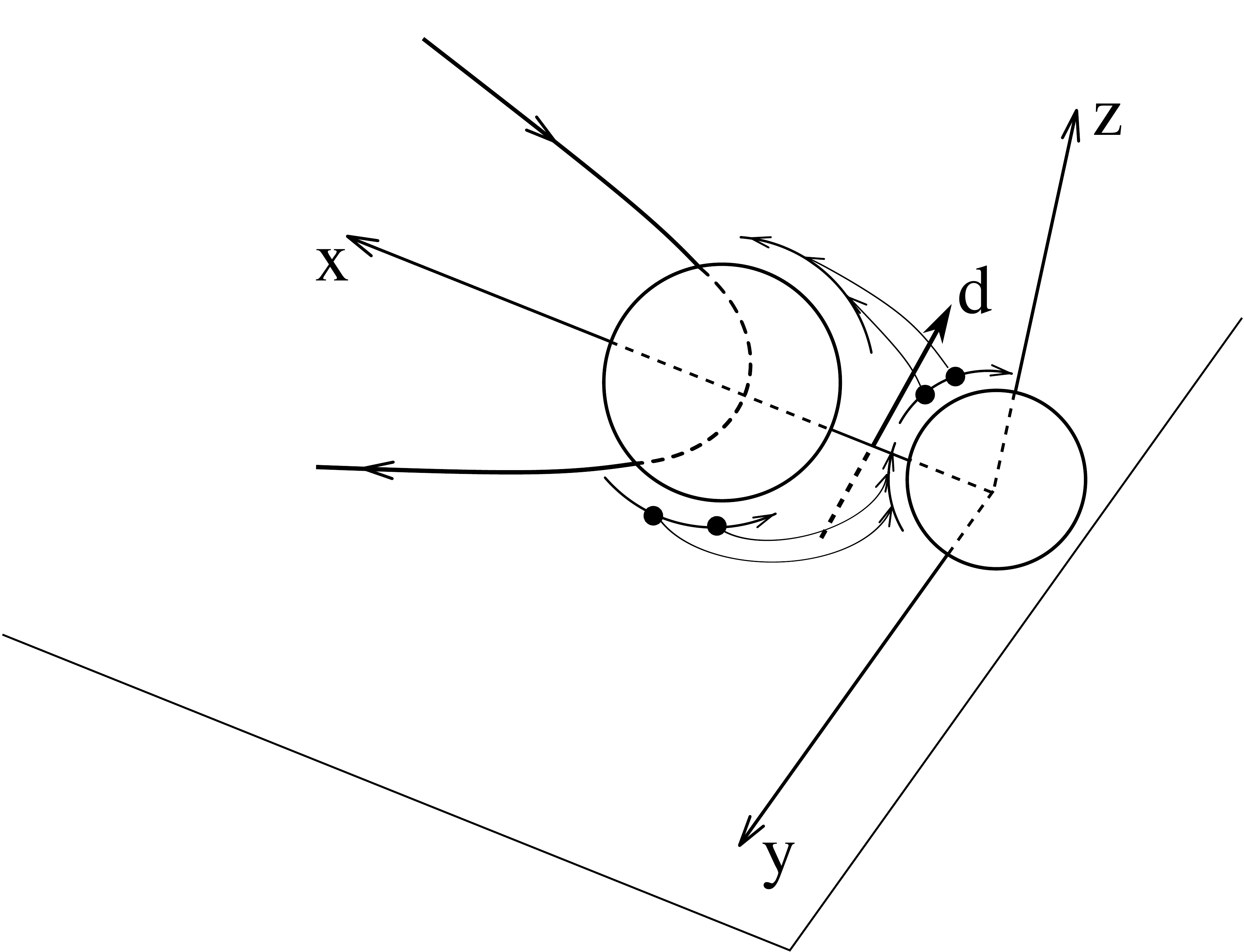}}
	\caption{Schematic representation of the quasielastic process in which a Cooper pair is transferred back and forth between two superconducting nuclei (e.g. projectile, $B(=A+2)\equiv^{116}$Sn and target, $b\equiv^{60}$Ni), i.e. $B+b\to F(=A+1)+f(=b+1)\to A+a(=b+2)\to F+f\to B+b$, in keeping with the fact that Cooper pair transfer is dominated by successive transfer. Because the transferred neutrons (black dots) carry an effective charge, an oscillating dipole with frequency $\nu=Q_{2n}/h$ is established, where $Q_{2n}$ is the $Q$-value of the reaction $^{116}$Sn+$^{60}$Ni$\to^{114}$Sn+$^{62}$Ni ($Q_{2n}$=1.307 MeV), and $\gamma$-rays are emitted by a time-dependent $\mathbf d$ dipole moment oscillating   in a plane which forms an angle of $63^\circ$ with   the reaction plane.}\label{fig:5}
\end{figure}
The neutron Cooper pair carries an effective charge $2\times e_{eff}=-e\times 2\left(78/176\right)\approx-e\times0.89$. Consequently,  the nuclear junction can be viewed as  biased by a potential $V=\left( Q_{2n}/(2\times e_{eff})\right)\approx-1.469$ MV, and the few cycles of the process (\ref{eq:12}) considered as an alternating, single Cooper pair, nuclear Josephson supercurrent of frequency $\nu_J=2\times e_{eff}V/h=Q_{2n}/h$.  As such,  and in keeping with the fact that the system is few MeV below the Coulomb barrier, where tunneling proceeds essentially  free of dissipation, one expects to be a source of ZHz photons emitted by a time-dependent dipole moment oscillating in a plane which forms an angle $\approx 63^\circ$ with the reaction plane, defined by the reaction condition of a smooth matching around the distance of closest approach of the projectile (Sn) neutron orbitals with that of the target (Ni) one (see Fig. \ref{fig:5}; see also p. 316 and Fig. 8, p. 326 of \cite{Broglia:04a}).

\begin{figure}
	\centerline{\includegraphics*[width=7cm,angle=0]{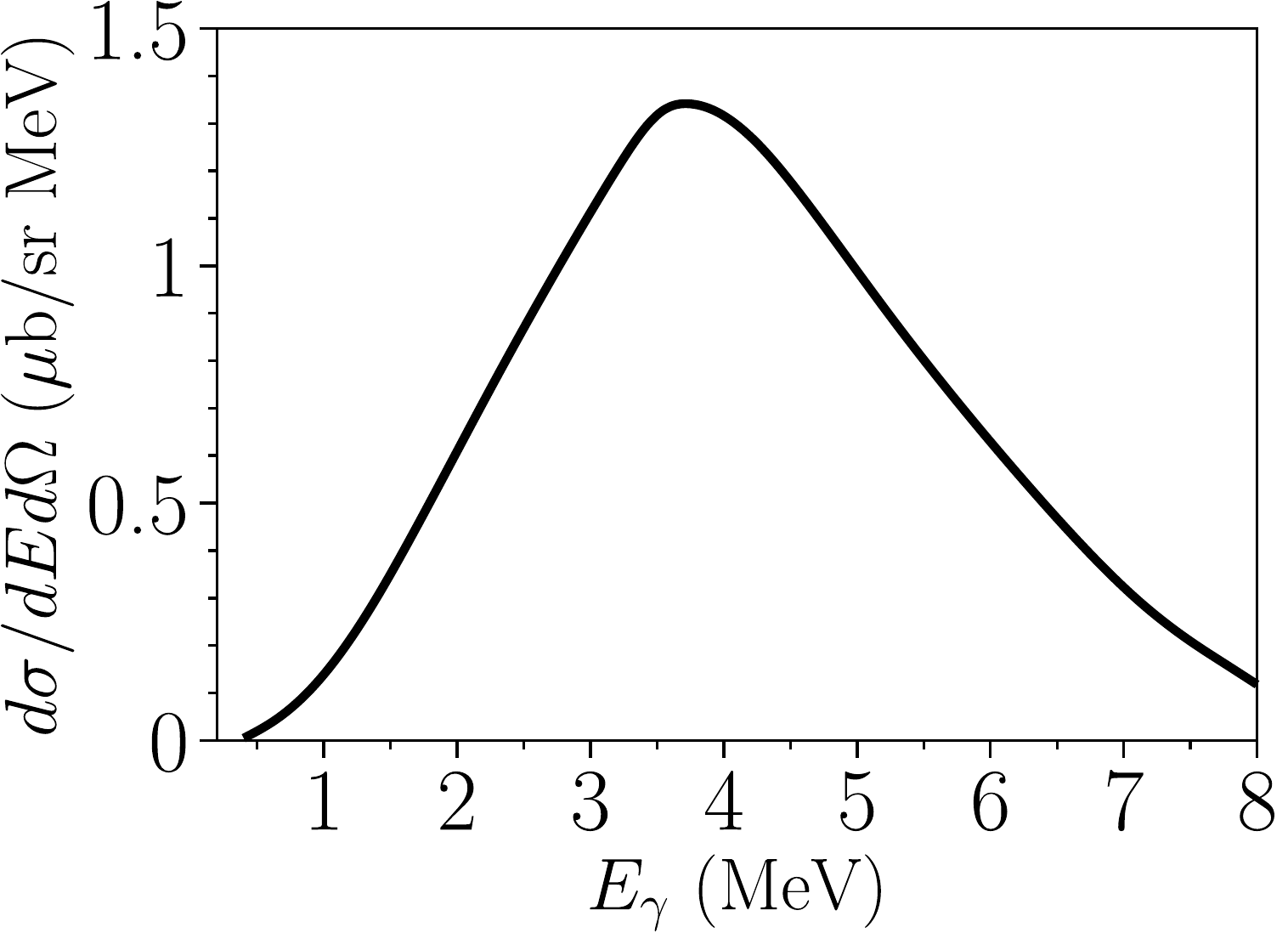}}
	\caption{Double differential cross section for $\gamma$-emission at $\theta_{c.m.}=140^\circ$ as a function of the energy of the emitted $\gamma$-ray, calculated with Eq. (\ref{eq:13}).}\label{fig:4}
\end{figure}

 In Fig. \ref{fig:4}, the $\gamma$-strength function
\begin{align}
  \label{eq:13}
 \nonumber \frac{d^2\sigma}{d\Omega dE_\gamma}&=\left(\frac{\mu_i\mu_f}{(2\pi\hbar^2)^2}\frac{k_f}{k_i}\right)\left(\frac{8\pi}{3}\frac{E^2_\gamma}{(\hbar c)^3}\right)\\
  &\times\left|T_{m_\gamma}(\mathbf k_f,\mathbf k_i)\right|^2\,\delta(E_\gamma+E_f-(E_i+Q_{2n})),
\end{align}
calculated as done in connection with $d\sigma_{2n}/d\Omega$, but introducing in the $T$-matrix also the dipole moment $d^1_{m_\gamma}=2\times e_{eff}\sqrt{\frac{4\pi}{3}}r Y^1_{m_\gamma}(\hat r)$, is shown.  This prediction, normalized to the angle integrated two nucleon transfer cross section (Fig. \ref{fig:2x}),   and the $\gamma$-phase factor provides  a quantum mechanical estimate of the dipole moment $\langle d\rangle\approx 0.89\times e\times r=e\times 9.36$ fm  of $r\approx10.52$ fm (correlation length). In the above relation $\langle d\rangle$ stands for the incoherent summation of the three contributions $m_\gamma=\pm1,0$. The back and forth transfer associated with the quasielastic process (\ref{eq:12}) is the nuclear analogue of an alternating Josephson current of a single Cooper pair expected to last for only few periods \cite{Potel:21}. It is of note that the dipole coupling to the electromagnetic field seems natural when one is dealing with electrons, while tunneling of neutrons, but also within the present context of protons, are not understood as the motion of (effective), real charged particles.

In keeping with the fact that one is able to predict absolute two-nucleon transfer differential cross sections involving superfluid nuclei within a 10\% accuracy \cite{Potel:13}, one expects the prediction displayed in Fig. \ref{fig:4} to be representative of heavy ion reactions between superfluid nuclei at energies below that of the Coulomb barrier

From the  results displayed in Fig. \ref{fig:2x} (see also Fig.\ref{fig:2}), and the predictions shown in Fig. \ref{fig:4}, one can ascribe the value of $\xi=12.0\pm1.5$ fm to the nuclear correlation length associated with  Cooper pair tunneling in  the (S-S) process  (\ref{eq:1}) (b). At the basis of the conspicuous uncertainty ascribed to the value of $\xi$ one finds a technical reason, namely the different sources of the two inputs ($D_0^{(c)}$ and dipole moment value), and a general physical one, namely the role played by pairing vibrations in finite quantum many-body systems as compared to condensed matter. Within this context we refer to studies of pairing phase transitions in superconducting metallic particles. In particular, in an ensemble of small Sn particles of radius $R\lesssim370$ \AA, deposited in vacuum and insulated from each other by oxide layers (\cite{Tsuboi:77,Muhlschlegel:72,Buhrman:73,Lauritzen:93,Perenboom:81}), which testify to the fact that while in bulk superconductors, the large intrinsic range of the pair coherence length implies a very narrow critical region (around $T_c)$, the size of this region increases as the dimension of the system decreases below the coherence length, leading essentially to zero-dimensional systems. A situation also found in the case of the BCS pairing phase transition observed in superfluid nuclei as a function of the rotational frequency (see e.g. \cite{Shimizu:89} and references therein). The critical frequency $(\omega_{rot})_{c}$ playing a role similar to that of the critical magnetic field $H_c$ in superconductors. 
\section{Conclusions}
BCS condensation of nuclear Cooper pairs leading to superfluid nuclei violates pair of particles conservation, thus defining a privileged direction in gauge space. The nuclear tweezer specific to tether the associated gauge angle cannot be but another system which also violates pairs of particles conservation. That is, another nuclear condensate in weak contact with the first one through the Josephson-like junction transiently established in a heavy-ion collision between superfluid nuclei. The associated single Cooper pair ``current'' being the specific probe of the intrinsic structure of the transferred Cooper pair. The corresponding absolute cross section normalized with respect to the single-quasiparticle transfer one, constitutes the connection to the experimental results where from one can read the value of $\xi$. One can then conclude that the mean square radius of nuclear Cooper pairs joins the ranks of those quantities characterizing the atomic nucleus which, in principle, are measurable.

Within this context one can  state that an important page in the understanding of pairing in nuclei has been written at the Laboratori Nazionale di Legnaro and the Dipartimento di Fisica dell'Universit\`a di Torino, which also provides  elements to attempt at relating  special effects observed in the study of superconductivity in metals, with similar processes associated with superconductivity in finite quantum many-body systems, of which the atomic nucleus can be viewed as a paradigm.

  This work was performed under the auspices of the U.S. Department of Energy by Lawrence Livermore National Laboratory under Contract No. DE-AC52-07NA27344. F. B. thanks the Spanish Ministerio de Econom\'ia y Competitividad and FEDER funds under project FIS2017-88410-P. This work is part of the I+D+i project with Ref. PID2020-114687GB-I00, funded by MCIN/AEI/10.13039/501100011033.




\begin{thebibliography}{10}

\bibitem{Heisenberg:25bis}
W.~Heisenberg.
\newblock {\"{U}ber} quantentheoretische umdeutung kinematischer und
  mechanischer beziehungen.
\newblock {\em {Z. Phys.}}, 33:879, 1925.
\newblock {English translation in B. L. Van der Waerden, Sources of Quantum
  Mechanics, Dover, New York 1968, p. 261}.

\bibitem{Bohr:58}
A.~Bohr, B.~R. Mottelson, and D.~Pines.
\newblock Possible analogy between the excitation spectra of nuclei and those
  of the superconducting metallic state.
\newblock {\em Physical Review}, 110:936, 1958.

\bibitem{Broglia:13}
R.~A. Broglia and V.~Zelevinsky, editors.
\newblock {\em {50 Years of Nuclear BCS}}.
\newblock World Scientific, Singapore, 2013.

\bibitem{Schilpp:51}
N.~Bohr.
\newblock {Discussion with Einstein on epistemological problems in atomic
  physics}.
\newblock In P.~A. Schilpp, editor, {\em Albert Einstein,
  Philosopher-Scientist}, page 199. Harper, New York, 1951.

\bibitem{Cooper:92}
L.~Cooper.
\newblock {Microscopic Quantum Interference Effects in the Theory of
  Superconductivity}.
\newblock In Stig Lundqvist, editor, {\em Nobel Lectures. Physics 1971-1980},
  page~73. World Scientific, Singapore, 1992.

\bibitem{Bardeen:92}
J.~Bardeen.
\newblock {Electron-Phonon Interactions and Superconductivity}.
\newblock In Stig Lundqvist, editor, {\em Nobel Lectures. Physics 1971-1980},
  page~54. World Scientific, Singapore, 1992.

\bibitem{Frohlich:52}
H.~Fr\"{o}hlich.
\newblock Interaction of electrons with lattice vibrations.
\newblock {\em Procs. Royal Soc.}, 215:291, 1952.

\bibitem{Schrieffer:92b}
J.~R. Schrieffer.
\newblock {John Bardeen and the theory of superconductivity.}
\newblock {\em Physics Today}, 45:46, 1992.

\bibitem{Schrieffer:92}
J.~R. Schrieffer.
\newblock {Macroscopic Quantum Phenomena from Pairing in Superconductors}.
\newblock In Stig Lundqvist, editor, {\em Nobel Lectures. Physics 1971-1980},
  page~97. World Scientific, Singapore, 1992.

\bibitem{Bardeen:57a}
J.~Bardeen, L.~N. Cooper, and J.~R. Schrieffer.
\newblock Microscopic theory of superconductivity.
\newblock {\em Physical Review}, 106:162, 1957.

\bibitem{Bardeen:57b}
J.~Bardeen, L.~N. Cooper, and J.~R. Schrieffer.
\newblock Theory of superconductivity.
\newblock {\em Physical Review}, 108:1175, 1957.

\bibitem{Penrose:51}
O.~Penrose.
\newblock {Bose-Einstein Condensation and Liquid Helium}.
\newblock {\em Philosophical Magazine}, 42:1373, 1951.

\bibitem{Penrose:56}
O.~Penrose and L.~Onsager.
\newblock {Bose-Einstein Condensation and Liquid Helium}.
\newblock {\em Phys. Rev.}, 104:576, 1956.

\bibitem{Anderson:96}
P.~W. Anderson.
\newblock {Off-diagonal long-range order and flux quantization}.
\newblock In H.~Holden and S.~Kjellstrup~Ratkje, editors, {\em {The collected
  works of Lars Onsager}}, page 729. World Scientific, Singapore, 1996.

\bibitem{Yang:62}
C.~N. Yang.
\newblock {Concept of Off-Diagonal Long-Range Order and the Quantum Phases of
  Liquid He and of Superconductors}.
\newblock {\em Rev. Mod. Phys.}, 34:694, 1962.

\bibitem{Schrieffer:64}
J.~R. Schrieffer.
\newblock {\em Superconductivity}.
\newblock Benjamin, New York, 1964.

\bibitem{Ginzburg:04}
Vitaly~L. Ginzburg.
\newblock {Nobel Lecture: On superconductivity and superfluidity (what I have
  and have not managed to do) as well as on the ``physical minimum'' at the
  beginning of the XXI century}.
\newblock {\em Rev. Mod. Phys.}, 76:981, 2004.

\bibitem{Bohr:76}
A.~Bohr.
\newblock {\em Rotational Motion in Nuclei, in Les Prix Nobel en 1975}.
\newblock Imprimerie Royale Norstedts Tryckeri, Stockholm, 1976.
\newblock p. 59.

\bibitem{Bes:66}
D.~R. B{\`{e}}s and R.~A. Broglia.
\newblock Pairing vibrations.
\newblock {\em Nucl. Phys.}, 80:289, 1966.

\bibitem{Brink:05}
D.~M. Brink and R.~A. Broglia.
\newblock {\em Nuclear Superfluidity}.
\newblock Cambridge University Press, Cambridge, 2005.

\bibitem{Broglia:00}
R.~A. Broglia, J.~Terasaki, and N.~Giovanardi.
\newblock The {A}nderson--{G}oldstone--{N}ambu mode in finite and in infinite
  systems.
\newblock {\em Physics Reports}, 335:1, 2000.

\bibitem{Hinohara:16}
Nobuo Hinohara and Witold Nazarewicz.
\newblock {Pairing Nambu-Goldstone Modes within Nuclear Density Functional
  Theory}.
\newblock {\em Phys. Rev. Lett.}, 116:152502, 2016.

\bibitem{Anderson:66}
P.~W. Anderson.
\newblock Considerations on the flow of superfluid helium.
\newblock {\em Rev. Mod. Phys.}, 38:298--310, 1966.

\bibitem{Barranco:99}
F.~Barranco, R.~A. Broglia, G.~Gori, E.~Vigezzi, P.~F. Bortignon, and
  J.~Terasaki.
\newblock Surface vibrations and the pairing interaction in nuclei.
\newblock {\em Phys. Rev. Lett.}, 83:2147, 1999.

\bibitem{Saperstein:12}
E.~E. Saperstein and M.~Baldo.
\newblock {Microscopic Origin of Pairing}.
\newblock In R.~A. Broglia and V.~Zelevinsky, editors, {\em 50 Years of Nuclear
  BCS}, page 263. World Scientific, Singapore, 2013.

\bibitem{Avdenkov:12}
A.~Avdeenkov and S.~Kamerdzhiev.
\newblock {Phonon Coupling and the Single-Particle Characteristics of Sn
  Isotopes}.
\newblock In R.~A. Broglia and V.~Zelevinsky, editors, {\em 50 Years of Nuclear
  BCS}, page 274. World Scientific, Singapore, 2013.

\bibitem{Lombardo:12}
U.~Lombardo, H.~J. Schulze, and W.~Zuo.
\newblock {Induced Pairing Interaction in Neutron Star Matter}.
\newblock In R.~A. Broglia and V.~Zelevinsky, editors, {\em 50 Years of Nuclear
  BCS}, page 338. World Scientific, Singapore, 2013.

\bibitem{Idini:15}
A.~Idini, G.~Potel, F.~Barranco, E.~Vigezzi, and R.~A. Broglia.
\newblock Interweaving of elementary modes of excitation in superfluid nuclei
  through particle-vibration coupling: Quantitative account of the variety of
  nuclear structure observables.
\newblock {\em Phys. Rev. C}, 92:031304, 2015.

\bibitem{Barranco:05}
F.~Barranco, P.~F. Bortignon, R.~A. Broglia, G.~Col\`o, P.~Schuck, E.~Vigezzi,
  and X.~Vi\~nas.
\newblock Pairing matrix elements and pairing gaps with bare, effective, and
  induced interactions.
\newblock {\em Phys. Rev. C}, 72:054314, 2005.

\bibitem{Bohr:69}
A.~Bohr and B.~R. Mottelson.
\newblock {\em Nuclear Structure, Vol.I}.
\newblock Benjamin, New York, 1969.

\bibitem{Bohr:75}
A.~Bohr and B.~R. Mottelson.
\newblock {\em Nuclear Structure, Vol.II}.
\newblock Benjamin, New York, 1975.

\bibitem{Mottelson:76}
B.~R. Mottelson.
\newblock {\em Elementary Modes of Excitation in Nuclei, Le Prix Nobel en
  1975}, page~80.
\newblock Imprimerie Royale Norstedts Tryckeri, Stockholm, 1976.

\bibitem{Mayer:48}
Maria~G. Mayer.
\newblock On closed shells in nuclei.
\newblock {\em Phys. Rev.}, 74:235, 1948.

\bibitem{Mayer:49}
Maria~Goeppert Mayer.
\newblock On closed shells in nuclei. {II}.
\newblock {\em Phys. Rev.}, 75:1969, 1949.

\bibitem{Mayer:55}
M.~G. Mayer and J.~Jensen.
\newblock {\em {Elementary Theory of Nuclear Structure}}.
\newblock Wiley, New York,NY, 1955.

\bibitem{Belyaev:59}
S.~T. Belyaev.
\newblock Effect of pairing correlations on nuclear properties.
\newblock {\em Kgl. Danske Videnskab. Selskab, Mat.-fys. Medd.}, 31:no. 11,
  1959.

\bibitem{Belyaev:13}
S.~T. Belyaev.
\newblock Pair correlations in nuclei: Copenhagen 1958.
\newblock In R.~A. Broglia and V.~Zelevinski, editors, {\em 50 Years of Nuclear
  BCS}, page~3. World Scientific, Singapore, 2013.

\bibitem{Bohr:64}
A.~Bohr.
\newblock Elementary modes of excitation and their coupling.
\newblock In {\em {Comptes Rendus du Congr\`{e}s International de Physique
  Nucl\'{e}aire}}, volume~1, page 487. Centre National de la Recherche
  Scientifique, 1964.

\bibitem{Hogassen:61}
J.~H\"ogaasen-Feldman.
\newblock A study of some approximations of the pairing force.
\newblock {\em Nuclear Physics}, 28:258, 1961.

\bibitem{Bjerregaard:66b}
J.~H. Bjerregaard, O.~Hansen, O.~Nathan, and S.~Hinds.
\newblock States of {$^{208}$Pb} from double triton stripping.
\newblock {\em {Nucl. Phys.}}, 89:337, 1966.

\bibitem{Broglia:67}
R.~A. Broglia and C.~Riedel.
\newblock Pairing vibration and particle-hole states excited in the reaction
  $^{206}${P}b(t, p)$^{208}${P}b.
\newblock {\em Nucl. Phys. A}, 92:145, 1967.

\bibitem{Potel:13b}
G.~Potel, A.~Idini, F.~Barranco, E.~Vigezzi, and R.~A. Broglia.
\newblock {Quantitative study of coherent pairing modes with two--neutron
  transfer: Sn isotopes}.
\newblock {\em {Phys. Rev. C}}, 87:054321, 2013.

\bibitem{Bracco:96}
A.~Bracco, F.~Alasia, S.~Leoni, E.~Vigezzi, and R.~A. Broglia.
\newblock Nuclear rotations.
\newblock {\em Contemporary Physics}, 37:183, 1996.

\bibitem{Dheer:61}
P.~N. Dheer.
\newblock The surface impedance of normal and superconducting indium at 3000
  mc/s.
\newblock {\em Proceedings of the Royal Society of London A}, 260:333, 1961.

\bibitem{Tilley:90}
D.~R. Tilley and J.~Tilley.
\newblock {\em Superfluidity and superconductivity}.
\newblock IOP, Bristol, 1990.

\bibitem{Chambers:52}
R.~G. Chambers.
\newblock The anomalous skin effect.
\newblock {\em Proceedings of the Royal Society of London A}, 215:481, 1952.

\bibitem{Pippard:53}
A.~B. Pippard.
\newblock An experimental and theoretical study of the relation between
  magnetic field and current in a superconductor.
\newblock {\em Proceedings of the Royal Society A}, 216:547, 1953.

\bibitem{London:35}
F.~London and H.~London.
\newblock The electromagnetic equations of the supraconductor.
\newblock {\em Proceedings of the Royal Society of London A}, 149:71, 1935.

\bibitem{London:54}
F.~London.
\newblock {\em {Superfluids}}.
\newblock John Wiley and Sons, New York, 1954.

\bibitem{Mersevey:69}
R~Mersevey and B.~B. Schwartz.
\newblock {Equilibrium properties: comparison of experimental results with
  prediction of BCS theory}.
\newblock In R.D. Parks, editor, {\em Superconductivity}, volume~1, page 117,
  New York, 1969. Marcel Dekker, Inc.

\bibitem{Pippard:12}
A.~B. Pippard.
\newblock {The historical context of Josephson discovery}.
\newblock In H.~Rogalla and P.~H. Kes, editors, {\em 100 years of
  superconductivity}, page~30. CRC Press, Taylor and Francis, FL, 2012.

\bibitem{Giaver:73}
I.~Giaever.
\newblock Electron tunneling and superconductivity.
\newblock In {\em Le Prix Nobel en 1973}, page~84. Norstedt, P.A. and
  S{\"{o}}ner, 1973.

\bibitem{Anderson:64b}
P.~W. Anderson.
\newblock Special effects in superconductivity.
\newblock In E.~R. Caianiello, editor, {\em The Many-Body Problem, Vol.2}, page
  113. Academic Press, New York, 1964.

\bibitem{Josephson:62}
B.~D. Josephson.
\newblock Possible new effects in superconductive tunnelling.
\newblock {\em Phys. Lett.}, 1:251, 1962.

\bibitem{Josephson:73}
B.~D. Josephson.
\newblock The discovery of tunneling supercurrents.
\newblock In P.A. Norstedt and S{\"{o}}ner, editors, {\em Le Prix Nobel en
  1973}, page 104. 1973.

\bibitem{Anderson:63}
P.~W. Anderson and J.~M. Rowell.
\newblock {Probable observation of the Josephson superconducting tunneling
  effect}.
\newblock {\em Physical Review Letters}, 10:230, 1963.

\bibitem{Shapiro:63}
S.~Shapiro.
\newblock Josephson currents in superconducting tunneling: The effect of
  microwaves and other observations.
\newblock {\em Phys. Rev. Lett.}, 11:80, 1963.

\bibitem{Lindelof:81}
P.~E. Lindelof.
\newblock {Superconducting micro bridges exhibiting Josephson properties}.
\newblock {\em Rep. Prog. Phys.}, 44:60, 1981.

\bibitem{Rogalla:12}
H.~Rogalla and P.~H. Kes, editors.
\newblock {\em 100 years of superconductivity}.
\newblock CRC Press, Taylor and Francis, FL, 2012.

\bibitem{Schrodinger:35}
E.~{Schr\"odinger}.
\newblock Discussion of probability relations between separated systems.
\newblock {\em Mathematical Proceedings of the Cambridge Philosophical
  Society}, 31:555, 1935.

\bibitem{Mottelson:02}
B.~Mottelson.
\newblock Elementary features of nuclear structure.
\newblock In H.~Nifenecker, J.~P. Blaizot, G.~F. Bertsch, W.~Weise, and
  F.~David, editors, {\em {Trends in Nuclear Physics, 100 years later, Les
  Houches, Session LXVI}}, page~25. Elsevier, Amsterdam, 1998.

\bibitem{Potel:17}
G.~Potel, A.~Idini, F.~Barranco, E.~Vigezzi, and R.~A. Broglia.
\newblock {From bare to renormalized order parameter in gauge space: structure
  and reactions}.
\newblock {\em Phys. Rev C}, 96:034606, 2017.

\bibitem{deBoer:48b}
J.~de~Boer and R.~J. Lundbeck.
\newblock {Quantum theory of condensed permanent gases III. The equation of
  state of liquids}.
\newblock {\em Physica}, 14:520, 1948.

\bibitem{deBoer:48}
J.~de~Boer.
\newblock Quantum theory of condensed permanent gases in the law of
  corresponding states.
\newblock {\em Physica}, 14:139, 1948.

\bibitem{deBoer:57}
J.~de~Boer.
\newblock {Quantum Effects and Exchange Effects on the Thermodynamic Properties
  of Liquid Helium}.
\newblock volume~2 of {\em Progress in Low Temperature Physics}, page~1. 1957.

\bibitem{Nosanow:76}
L.H. Nosanow.
\newblock {On the possible superfluidity of $^6$He--Its phase diagram and those
  of $^6$He-$^4$He and $^6$He-$^3$He mixtures}.
\newblock {\em Journal of Low Temperature Physics}, 23:605, 1976.

\bibitem{Anderson:84}
P.~W. Anderson and D.~L. Stein.
\newblock {Broken symmetry, emergent properties, dissipative structures, life:
  are they related?}
\newblock In {\em {P. W. Anderson}, Basic notions of condensed matter}, page
  263. Benjamin, Menlo Park, CA, 1984.

\bibitem{Montanari:14}
D.~Montanari, L.~Corradi, S.~Szilner, G.~Pollarolo, E.~Fioretto, G.~Montagnoli,
  F.~Scarlassara, A.~M. Stefanini, S.~Courtin, A.~Goasduff, F.~Haas,
  D.~Jelavi\ifmmode \acute{c}\else~\'{c}\fi{} Malenica, C.~Michelagnoli,
  T.~Mijatovi\ifmmode~\acute{c}\else \'{c}\fi{}, N.~Soi\ifmmode~\acute{c}\else
  \'{c}\fi{}, C.~A. Ur, and M.~Varga~Pajtler.
\newblock {Neutron Pair Transfer in $^{60}\mathrm{Ni}+^{116}\mathrm{Sn}$ Far
  below the Coulomb Barrier}.
\newblock {\em Phys. Rev. Lett.}, 113:052501, 2014.

\bibitem{Montanari:16}
D.~Montanari, L.~Corradi, S.~Szilner, G.~Pollarolo, A.~Goasduff,
  T.~Mijatovi\ifmmode~\acute{c}\else \'{c}\fi{}, D.~Bazzacco, B.~Birkenbach,
  A.~Bracco, L.~Charles, S.~Courtin, P.~D\'esesquelles, E.~Fioretto, A.~Gadea,
  A.~G\"orgen, A.~Gottardo, J.~Grebosz, F.~Haas, H.~Hess, D.~Jelavi\ifmmode
  \acute{c}\else~\'{c}\fi{} Malenica, A.~Jungclaus, M.~Karolak, S.~Leoni,
  A.~Maj, R.~Menegazzo, D.~Mengoni, C.~Michelagnoli, G.~Montagnoli, D.~R.
  Napoli, A.~Pullia, F.~Recchia, P.~Reiter, D.~Rosso, M.~D. Salsac,
  F.~Scarlassara, P.-A. S\"oderstr\"om, N.~Soi\ifmmode~\acute{c}\else
  \'{c}\fi{}, A.~M. Stefanini, O.~Stezowski, Ch. Theisen, C.~A. Ur, J.~J.
  Valiente-Dob\'on, and M.~Varga~Pajtler.
\newblock Pair neutron transfer in
  $^{60}\text{Ni}+\phantom{\rule{0.16em}{0ex}}^{116}\text{Sn}$ probed via
  $\ensuremath{\gamma}$-particle coincidences.
\newblock {\em Phys. Rev. C}, 93:054623, 2016.

\bibitem{Magierski:17}
P.~Magierski, K.~Sekizawa, and G.~Wlaz\l{}owski.
\newblock Novel role of superfluidity in low-energy nuclear reactions.
\newblock {\em Phys. Rev. Lett.}, 119:042501, 2017.

\bibitem{Magierski:17b}
P.~Magierski, K.~Sekizawa, and G.~Wlaz\l{}owski.
\newblock Novel role of superfluidity in low-energy nuclear reactions.
\newblock {\em Phys. Rev. Lett.}, 119:042501, 2017.
\newblock Supplemental material.

\bibitem{Lee:09}
J.~Lee, M.~B. Tsang, W.~G. Lynch, M.~Horoi, and S.~C. Su.
\newblock {Neutron spectroscopic factors of Ni isotopes from transfer
  reactions}.
\newblock {\em Phys. Rev. C}, 79:054611, 2009.

\bibitem{Satchler:80}
G.R. Satchler.
\newblock {\em Introduction to Nuclear Reactions}.
\newblock Mc Millan, New York, 1980.

\bibitem{Broglia:81b}
R.~A. Broglia, G.~Pollarolo, and A.~Winther.
\newblock On the absorptive potential in heavy ion scattering.
\newblock {\em Nuclear Physics A}, 361:307, 1981.

\bibitem{Pollarolo:83}
G.~Pollarolo, R.~A. Broglia, and A.~Winther.
\newblock Calculation of the imaginary part of the heavy ion potential.
\newblock {\em Nuclear Physics A}, 406:369, 1983.

\bibitem{Broglia:04a}
R.~A. Broglia and A.~Winther.
\newblock {\em Heavy Ion Reactions}.
\newblock Westview Press, Boulder, CO., 2004.

\bibitem{Sorensen:92b}
J.H. S{\o{}}rensen and A.~Winther.
\newblock The absorptive potential for heavy-ion collisions at intermediate and
  low energy.
\newblock {\em Nuclear Physics A}, 550:329, 1992.

\bibitem{Potel:21}
G.~Potel, F.~Barranco, E.~Vigezzi, and R.~A. Broglia.
\newblock {Quantum entanglement in nuclear Cooper-pair tunneling with
  $\ensuremath{\gamma}$ rays}.
\newblock {\em Phys. Rev. C}, 103:L021601, 2021.

\bibitem{Broglia:21}
R.~A. Broglia, F.~Barranco, G.~Potel, and E.~Vigezzi.
\newblock {Transient Weak Links between Superconducting Nuclei: Coherence
  Length}.
\newblock {\em Nuclear Physics News}, 31, No 4:24, 2021.

\bibitem{Magierski:21}
P.~Magierski.
\newblock {The Tiniest Superfluid Circuit in Nature}.
\newblock {\em Physics}, 14:27, 2021.

\bibitem{Potel:13}
G.~Potel, A.~Idini, F.~Barranco, E.~Vigezzi, and R.~A. Broglia.
\newblock Cooper pair transfer in nuclei.
\newblock {\em {Rep. Prog. Phys.}}, 76:106301, 2013.

\bibitem{Tsuboi:77}
T.~Tsuboi and T.~Suzuki.
\newblock {Specific Heat of Superconducting Fine Particles of Tin. I.
  Fluctuations in Zero Magnetic Field}.
\newblock {\em Journal of the Physical Society of Japan}, 42:437, 1977.

\bibitem{Muhlschlegel:72}
B.~{M\"uhlschlegel}, D.~J. Scalapino, and R.~Denton.
\newblock Thermodynamic properties of small superconducting particles.
\newblock {\em Phys. Rev. B}, 6:1767, 1972.

\bibitem{Buhrman:73}
R.~A. Buhrman and W.~P. Halperin.
\newblock {Fluctuation Diamagnetism in a {``Zero-Dimensional''}
  Superconductor}.
\newblock {\em Phys. Rev. Lett.}, 30:692, 1973.

\bibitem{Lauritzen:93}
B.~Lauritzen, A.~Anselmino, P.~F. Bortignon, and R.~A. Broglia.
\newblock Pairing phase transition in small particles.
\newblock {\em Annals of Physics}, 223:216, 1993.

\bibitem{Perenboom:81}
J.~A. A.~J. Perenboom, P.~Wyder, and F.~Meier.
\newblock Electronic properties of small metallic particles.
\newblock {\em Physics Reports}, 78:173, 1981.

\bibitem{Shimizu:89}
Y.~R. Shimizu, J.~D. Garrett, R.~A. Broglia, M.~Gallardo, and E.~Vigezzi.
\newblock Pairing fluctuations in rapidly rotating nuclei.
\newblock {\em Reviews of Modern Physics}, 61:131, 1989.

\end{thebibliography}
  \end{document}